\documentclass[twocolumn,pra,showpacs,preprintnumbers,amsmath,amssymb]{revtex4-1}
\usepackage{xcolor}
\usepackage{graphicx}
\usepackage{subfigure}  
\usepackage{hyperref} 
\usepackage{multirow}

\DeclareMathOperator{\Tr}{Tr}
\def\ket #1{\vert #1\rangle}
\def\bra #1{\langle #1\vert}
\newcommand{\ketbra}[2]{\ensuremath{\ket{#1}\!\bra{#2}}}

\newcommand{\Jam}{Jamio\l kowski }

\newcommand{\Zp}{\ensuremath{\mathbb{Z}_p}}

\newcommand{\Ua}{\ensuremath{U_{\upsilon}}}

\def\ket #1{\vert #1\rangle}
\def\bra #1{\langle #1\vert}
\def\ketbra #1#2{\ket{#1}\!\bra{#2}}

\newcommand{\w}{\ensuremath{\mathrm{\omega}}}

\renewcommand{\i}{\mathrm{i}}

\newcommand{\U}{\mathrm{U}}
\newcommand{\SU}{\mathrm{SU}}
\newcommand{\I}{\mathbb{I}}

\begin{document}
\title{Qudit versions of the qubit ``pi-over-eight'' gate}

\author{Mark Howard}
\email{mark.howard@nuim.ie}
\thanks{}
\affiliation{%
Department of Mathematical Physics, National University of Ireland, Maynooth, Ireland}
\author{Jiri Vala}
\email{Jiri.Vala@nuim.ie}
\affiliation{%
Department of Mathematical Physics, National University of Ireland, Maynooth, Ireland \\
Dublin Institute for Advanced Studies, School of Theoretical Physics, 10 Burlington Road, Dublin, Ireland
}%
\date{\today}

\begin{abstract} 
When visualised as an operation on the Bloch sphere, the qubit ``pi-over-eight'' gate corresponds to one-eighth of a complete rotation about the vertical axis. This simple gate often plays an important role in quantum information theory, typically in situations for which Pauli and Clifford gates are insufficient. Most notably, when it supplements the set of Clifford gates then universal quantum computation can be achieved.  The ``pi-over-eight'' gate is the simplest example of an operation from the third level of the Clifford hierarchy (i.e., it maps Pauli operations to Clifford operations under conjugation). Here we derive explicit expressions for all qudit (d-level, where d is prime) versions of this gate and analyze the resulting group structure that is generated by these diagonal gates. This group structure differs depending on whether the dimensionality of the  qudit is two, three or greater than three. We then discuss the geometrical relationship of these gates (and associated states) with respect to Clifford gates and stabilizer states. We present evidence that these gates are maximally robust to depolarizing and phase damping noise, in complete analogy with the qubit case. Motivated by this and other similarities we conjecture that these gates could be useful for the task of qudit magic-state distillation and, by extension, fault-tolerant quantum computing. Very recent, independent work by Campbell, Anwar and Browne confirms the correctness of this intuition, and we build upon their work to characterize noise regimes for which noisy implementations of these gates can (or provably cannot) supplement Clifford gates to enable universal quantum computation.
\end{abstract}

\maketitle

\section{Background and Motivation}
The qubit ``pi-over-eight'' gate, $U_{\pi/8}$,
\begin{align}
U_{\pi/8}=\left(\begin{array}{cc}
e^{-i\frac{\pi}{8}} & 0\\
0 & e^{i\frac{\pi}{8}}
\end{array}\right)
\end{align}
plays a special role in a number of quantum informational tasks. The Gottesman-Knill theorem \cite{Gottesman:1998} tells us that a circuit using only Clifford gates and Pauli measurements (i.e., a stabilizer circuit) is insufficient for universal quantum computation (UQC). Whilst technically, adding the ability to perform any single-qubit non-Clifford gate is sufficient for obtaining UQC, one typically sees that the $U_{\pi/8}$ is chosen as the most natural and easiest with which to work \cite{Boykin2000}. In measurement-based scenarios, supplementing Pauli measurement directions with an additional rotated (by $U_{\pi/8}$) measurement basis can also enable the performance of new tasks. For example, performing Pauli measurements on the Bell state $(\ket{00}+\ket{11})/\sqrt{2}$, or any two-qubit stabilizer state, does not exhibit any better-than-classical performance in a nonlocal CHSH game \cite{Buhrman2001,Howard2012}, whereas the introduction of the new rotated basis enables the optimum quantum advantage. Measurements in the aforementioned rotated basis appear to arise naturally in other quantum-informational tasks too e.g., universal blind quantum computation \cite{Broadbent:2009}, where Pauli measurements and operators would be insufficient.

Arguably, much of the utility of this gate arises from its close relationship with the Clifford group, whilst still not being a member of the group. In fact, $U_{\pi/8}$ is the simplest meaningful example of a gate from the third level of the Clifford hierarchy (defined later), the first two levels of which correspond to Pauli gates and Clifford gates. Operations from the Clifford hierarchy have properties that make them suitable for teleportation-based UQC \cite{Gottesman1999b}, transversal implementation (see below), learning an unknown gate \cite{Low2009}, or secure assisted quantum computation \cite{Childs2005}. Gates from higher levels of the Clifford hierarchy are also related to so-called semi-Clifford gates \cite{Beigi:2010,Zeng2008,Gross:2008}.

It is now known \cite{Eastin2009} that no quantum error-correcting code allows for transversal (i.e., bitwise and manifestly fault-tolerant) implementation of a universal set of gates. Typically, many stabilizer codes (e.g., CSS codes) enable transversal implementation of the complete set of Clifford gates, but only these gates. An intriguing exception is provided by the $[[15,1,3]]$ punctured Reed-Muller code that allows for transversal implementation of the $U_{\pi/8}$ gate, but not the complete set of Clifford gates. Zeng \emph{et al.} \cite{Zeng2007} provide quite a detailed discussion on the relationship between Reed-Muller codes and transversal gates from the third (or higher) level of the Clifford hierarchy. It appeared that the transversal $U_{\pi/8}$ property of the $[[15,1,3]]$ code was useful in the derivation of a magic state distillation (MSD) routine by Bravyi and Kitaev \cite{BravyiKitaev:2005}, a routine that iteratively distills increasing pure copies of the state $\ket{\psi_{U_{\pi/8}}}=U_{\pi/8}\ket{+}\propto \ket{0}+e^{i\frac{\pi}{4}}\ket{1}$.

Independently of the foregoing discussion, it is also notable that $U_{\pi/8}$ is remarkable due its geometrical relationship with the set of Clifford gates -- it is the single-qubit unitary $U \in \SU(2)$ that is farthest outside the convex hull of Clifford operations. The associated single-qubit state $\ket{\psi_{U_{\pi/8}}}$, mentioned previously, is also remarkable in its convex-geometrical relationship with Pauli eigenstates. Furthermore, these geometrical relationships have ramifications for the amount of noise that can be tolerated by imperfect implementation of $U_{\pi/8}$  or imperfect preparation of $\ket{\psi_{U_{\pi/8}}}$, a scenario that arises naturally in any fault-tolerant UQC proposal that uses magic state distillation.

The state $\ket{\psi_{U_{\pi/8}}}\propto \ket{0}+e^{i\frac{\pi}{4}}\ket{1}$ is already known in quantum information theory as $\ket{H}$ -- a qubit magic state as defined in \cite{Knill05,BravyiKitaev:2005}. In addition, the most non-stabilizer qubit state $\ket{T}$ \cite{BravyiKitaev:2005}, in a convex-geometrical sense, is also a magic state. Moreover, both $\ket{H}$ and $\ket{T}$ are eigenvectors of Clifford gates. The importance of geometrically significant states and gates to qubit-based fault-tolerant UQC provided the motivation in \cite{WvDMH:2010} to find the most robust qudit states (for all prime dimension) and qudit gates (for $p \in \{2,3,5,7\}$). The maximally robust states (analogous to $\ket{T}$) were also found to be eigenvectors of Clifford gates, whereas the maximally robust gates had a strikingly simple form, which prompted the question whether these gates were related to generalized versions of the $U_{\pi/8}$ gate. Here, starting from the condition that these gates must be diagonal elements of $\mathcal{C}_3$, we derive generalized versions of $U_{\pi/8}$, which we call $\Ua$ and show that these are identical, up to an unimportant factor of a Clifford gate, to the maximally robust gates found in \cite{WvDMH:2010}. We also show that the associated states $\ket{\psi_{\Ua}}=\Ua \ket{+}$ are eigenvectors of Clifford gates, and that they obey a similar relationship with respect to stabilizer states as $\ket{H}$ does.

As we completed this work we became aware of very recent results by Campbell, Anwar and Browne \cite{Campbell:arxiv12}. There, the authors prove the existence of magic state distillation protocols (MSD) for all (prime) qudit systems, wherein the non-stabilizer states that are distilled are states that we have called $\ket{\psi_{\Ua}}$ here. Moreover, they show that $\Ua$ have a transversal implementation in a family of qudit Reed-Muller codes and they explain why this property is useful in MSD. It is hoped that the results presented here might aid in the analysis of such qudit MSD protocols. More generally, it seems likely that the gates $\Ua$ will find application in other areas of quantum information theory, particularly in qudit generalizations of qubit-based tasks for which $U_{\pi/8}$ is known to be helpful.

We begin Sec.~\ref{sec:Basic Mathematical Structure} by deriving explicit expressions for all qudit $\Ua$, and then proceed to analyze the resulting group structure where we note an interesting difference depending on whether $p=2$, $p=3$ or $p>3$. In Sec.~\ref{sec:Geometrical Features} we discuss geometrical features of these gates $\Ua$ and associated qudit states $\ket{\psi_{\Ua}}$ with a particular eye to applications in quantum computation. We conclude in section Sec.~\ref{sec:Possible Applications} with some observations on noise thresholds for qudit-based quantum computation.


\section{Basic Mathematical Structure}\label{sec:Basic Mathematical Structure}
\subsection{Generalized Pauli and Clifford Groups}\label{sec:Generalized Pauli and Clifford Groups}
Throughout, we always assume the dimension $p$, of a single particle, to be a prime number. Generalized versions of the familiar $\sigma_x$ and $\sigma_z$ Pauli operators, are defined \cite{Gottesman1999a} for $p>2$ as
\begin{align}
X\ket{j}=\ket{{j+1}\bmod{p}} \quad Z\ket{j}=\omega^{j}\ket{j} \label{eqn:PauliDef}
\end{align}
where $\omega=e^{2\pi i/p}$ is a primitive $p$-th root of unity such that $XZ=\omega^{-1}ZX$. In general, products of these Pauli operators are often called displacement operators,
\begin{align}
&D_{(x|z)}=\tau^{x z} X^{x} Z^{z}\quad \tau=e^{(p+1)\pi\i/p}=\omega^{2^{-1}} \label{eqn:Dxzdef}
\end{align}
where the format of the subscript $(x|z)$ reminds us of the symplectic form (often used in calculations involving Pauli operators e.g., error-correcting codes). The Weyl-Heisenberg group (or generalized Pauli group) for a single qudit is given by
\begin{align}
\mathcal{G} =\left\{\tau^c D_{\vec{\chi}} \vert \vec{\chi} \in \Zp^2, c \in \Zp\right\} \quad \left(\Zp=\{0,1,\ldots,p-1\}\right),
\end{align}
where $\vec{\chi}$ is a $2$-vector with elements from $\Zp$,so that $|\mathcal{G}|=p^2$ in situations where global phases can be ignored. The set of unitary operators that map the Pauli group onto itself under conjugation is called the Clifford group, $\mathcal{C}$,
\begin{align*}
 \mathcal{C}=\{C\in U(p)\vert C \mathcal{G} C^\dag =\mathcal{G}\}.
\end{align*}
The number of distinct Clifford gates for a single qudit system (ignoring global phases) is $|\mathcal{C}|=p^3(p^2-1)$.

Gottesman and Chuang \cite{Gottesman1999b} introduced the so-called Clifford hierarchy, a recursively defined set of gates given by
\begin{align}
\mathcal{C}_{k+1}=\left\{U \vert U\mathcal{C}_{1}U^\dag \subseteq \mathcal{C}_{k}\right\}
\end{align}
where $\mathcal{C}_{1}$ is the Pauli group. One obtains nested sets of operators, the first two sets of which correspond to elements of the Pauli and Clifford groups respectively, i.e.,
\begin{align}
\mathcal{G} \subseteq \mathcal{C} \subseteq \mathcal{C}_3 \subseteq \ldots
\end{align}
or equivalently in their notation
\begin{align}
\mathcal{C}_1 \subseteq \mathcal{C}_2 \subseteq \mathcal{C}_3 \subseteq \ldots
\end{align}
It is known \cite{Gottesman1999b,Beigi:2010,Zeng2008} that $\mathcal{C}_3$ (and above) does not form a group, although the diagonal subset of $\mathcal{C}_3$, that we study here, does.

The complete set of Clifford unitaries $\mathcal{C} \subset \U(p)$ is covered by varying over all $F \in SL(2,\Zp)$ and $\vec{\chi} \in \Zp^2$,
\begin{align}
\mathcal{C}=\{ C_{(F\vert \vec{\chi})}\ \vert \ F \in SL(2,\Zp), \vec{\chi} \in \Zp^2 \}, \label{eqn:CliffIso}
\end{align}
where $SL(2,\Zp)$ is the group whose elements are $2 \times 2$ matrices with unit determinant and matrix elements from $\Zp$.
The explicit recipe \cite{Appleby:arxiv09} for constructing a Clifford unitary with $F=\bigl(\begin{smallmatrix} \alpha & \beta \\ \gamma & \delta \end{smallmatrix}\bigr), \vec{\chi}=\bigl(\begin{smallmatrix} x\\ z \end{smallmatrix}\bigr)$ is given by
\begin{align}
&C_{(F\vert \vec{\chi})}=D_{(x|z)}V_F \label{eqn:CeqDxzVF} \\
&V_F=\begin{cases}
\frac{1}{\sqrt{p}}\sum_{j,k=0}^{p-1}\tau^{\beta^{-1}\left(\alpha k^2-2 j k +\delta j^2\right)}\ket{j}\bra{k}\quad &\beta\neq0\\
\sum_{k=0}^{p-1} \tau^{\alpha \gamma k^2} \ket{\alpha k}\bra{k}\quad &\beta=0.
\end{cases}\nonumber
\end{align}
Note that when $F=\mathbb{I}_2$ we have $V_F=\mathbb{I}_p$. Also
\begin{align}
V_F D_{(x|z)} V_F^\dag = D_{(\alpha x+\beta z| \gamma x + \delta z)}\\
C_{\left(F_1 \mid \vec{\chi}_1 \right)}C_{\left(F_2 \mid \vec{\chi}_2 \right)}\propto C_{\left(F_1F_2 \mid \vec{\chi}_1 + F_1 \vec{\chi}_2 \right)}\label{eqn:semidirectproduct}
\end{align}
where the proportionality symbol denotes equality modulo a global phase.

The particular case $\beta=0$ of Eq.~\eqref{eqn:CeqDxzVF} turns out to be particularly relevant to our investigation, and so we note that
\begin{align}
\det \left(\sum_{k=0}^{p-1} \tau^{\alpha\gamma k^2} \ketbra{k}{k}\right) &= \tau^{\tfrac{\alpha\gamma}{6} (2p-1)(p-1)p},&\\
&=1 \quad \forall\ p>3,&\\
&=\tau^{2\alpha\gamma} \ p=3,&
\end{align}
which has ramifications for results in the next section. (This peculiarity, for $p=3$, was also noted by Zhu \cite{Zhu:2010} in the course of an investigation into Weyl-Heisenberg covariant SIC-POVMs.) In particular we will use 
\begin{align}
C_{\left(\bigl[\begin{smallmatrix} 1 & 0\\ \gamma & 1 \end{smallmatrix}\bigr]\middle\vert \bigl[\begin{smallmatrix} x \\z \end{smallmatrix}\bigr] \right) } \in \SU(p) \quad \forall p>3, \label{eqn:DetVF}\\
\det \left(C_{\left(\bigl[\begin{smallmatrix} 1 & 0\\ \gamma & 1 \end{smallmatrix}\bigr]\middle\vert \bigl[\begin{smallmatrix} x \\z \end{smallmatrix}\bigr] \right) }\right)=\tau^{2\gamma} \text{ for }p=3,\nonumber
\end{align}
which eventually leads to an unusual group structure for qutrit generalizations of $\U_{\pi/8}$.

\subsection{Explicit form of qudit gates analogous to $U_{\pi/8}$}\label{sec:Explicitformofquditgates}

For simplicity and, in analogy with the qubit $U_{\pi/8}$-gate, we chose $\Ua$ (our putative higher-dimensional generalizations of $U_{\pi/8}$) to be diagonal in the computational basis, so that $U D_{(0|1)} U^\dag = D_{(0|1)}$. We claim that, for $p>3$, $\Ua$ can be written in the following form:
\begin{align}
\Ua=U(\upsilon_0,\upsilon_1,\ldots)=\sum_{j=0}^{p-1} \omega^{\upsilon_k} \ketbra{k}{k} \quad (\upsilon_k \in \Zp),\label{eqn:Uadef}
\end{align}
where $\omega=e^{\frac{2 \pi i}{p}}$, as usual. Note that $\det(\Ua)=\omega^{\sum_{k=0}^{p-1} \upsilon_k}$ so that $\protect{\Ua \in \SU(p)}$ if $\sum_{k=0}^{p-1} \upsilon_k = 0 \pmod p $. Straightforward application of Eq.~\eqref{eqn:PauliDef}, \eqref{eqn:Dxzdef} and \eqref{eqn:Uadef} gives
\begin{align}
\Ua  D_{(x|z)} \Ua^\dag&=D_{(x|z)} \sum_k \omega^{(\upsilon_{k+x}- \upsilon_{k})}\ketbra{k}{k} \label{eqn:DxzWx}
\end{align}
If \Ua\ is to be a member of $\mathcal{C}_3$ we require the right hand side of \eqref{eqn:DxzWx} to be a Clifford gate. Since $\Ua D_{(0|z)} \Ua^\dag = D_{(0|z)}$, trivially, we focus on the case $\Ua D_{(1|0)} \Ua^\dag$ in order to derive explicit expressions for \Ua.

Define $\gamma^\prime,z^\prime,\epsilon^\prime \in \Zp$ such that
\begin{align}
&\Ua D_{(1|0)} \Ua^\dag = \omega^{\epsilon^\prime} C_{\left(\bigl[\begin{smallmatrix} 1 & 0\\ \gamma^\prime & 1 \end{smallmatrix}\bigr]\middle\vert \bigl[\begin{smallmatrix} 1 \\z^\prime\end{smallmatrix}\bigr] \right) } \label{eqn:UDUeqC}
\end{align}
The fact that the right hand side of the Eq.~\eqref{eqn:UDUeqC} is the most general form can be seen by reference to Eq.~\eqref{eqn:CeqDxzVF} and \eqref{eqn:DxzWx}, and also by noting that $U \in \SU(p)$ implies $\omega^k U \in \SU(p)$, for any integer $k$.

Note that the right hand side of Eq.~\eqref{eqn:UDUeqC} represents a Pauli operator if and only if $\gamma^\prime=0$. Consequently, $\Ua$ must, by definition, be a (diagonal) Clifford operation in those cases when $\gamma^\prime=0$.
%

To solve the matrix equation Eq.~\eqref{eqn:UDUeqC}, begin by substituting Eq.~\eqref{eqn:DxzWx} so that


\begin{align}
& D_{(1|0)} \sum_k \omega^{(\upsilon_{k+1}- \upsilon_{k})}\ketbra{k}{k}=\omega^{\epsilon^\prime}C_{\left(\bigl[\begin{smallmatrix} 1 & 0\\ \gamma^\prime & 1 \end{smallmatrix}\bigr]\middle\vert \bigl[\begin{smallmatrix} 1 \\z^\prime\end{smallmatrix}\bigr] \right) }\label{eqn:Matrixeqn}
\end{align}
and use Eq.~\eqref{eqn:CeqDxzVF} to obtain
\begin{align}
& D_{(1|0)} \sum_k \omega^{(\upsilon_{k+1}- \upsilon_{k})}\ketbra{k}{k}=\omega^{\epsilon^\prime}D_{(1|z^\prime)} \sum_{k=0}^{p-1} \tau^{\gamma^\prime k^2} \ket{ k}\bra{k}
\end{align}
After canceling common factors of $D_{(1|0)}$ one is left with an identity between two diagonal matrices, so that
\begin{align}
\omega^{\upsilon_{k+1}-\upsilon_k}=\omega^{\epsilon^\prime}\tau^{z^\prime}\omega^{kz^\prime}\tau^{k^2 \gamma^\prime} \quad (\forall k \in \Zp),
\end{align}
%
%
%
or equivalently, using $\tau=\omega^{2^{-1}}$,
\begin{align}
\upsilon_{k+1}-\upsilon_k=\epsilon^\prime + 2^{-1}z^\prime +kz^\prime +2^{-1}k^2 \gamma^\prime .
\end{align}
This gives the recurrence relation
\begin{align}
\upsilon_{k+1}=\upsilon_k+k(2^{-1}k\gamma^\prime+z^\prime) + 2^{-1}z^\prime+\epsilon^\prime.
\end{align}
With a boundary condition $\upsilon_0=0$, we can solve to obtain
\begin{align}
\upsilon_k=\frac{1}{12} k (\gamma^\prime +k (6 z^\prime+(2 k-3) \gamma^\prime ))+k {\epsilon^\prime}  \label{eqn:recurrence}
\end{align}
where factors like $12^{-1}$ are understood to be evaluated modulo $p$.

For example, with $p=5$ and choosing $z^\prime=1,\gamma^\prime=4$ and $\epsilon^\prime=0$, we get
\begin{align}
&\upsilon=(\upsilon_0,\upsilon_1,\upsilon_2,\upsilon_3,\upsilon_4)=(0,3,4,2,1)\\
\Rightarrow &\Ua(z^\prime,\gamma^\prime,\epsilon^\prime)=\left(
      \begin{array}{ccccc}
        1 & 0 & 0 & 0 & 0 \\
        0 & e^{-\frac{4 \pi \i}{5}} & 0 & 0 & 0 \\
        0 & 0 & e^{-\frac{2 \pi \i}{5}} & 0 & 0 \\
        0 & 0 & 0 & e^{\frac{4 \pi \i}{5}} & 0 \\
        0 & 0 & 0 & 0 & e^{\frac{2 \pi \i}{5}} \\
      \end{array}
    \right)\label{eqn:Uexample5}
\end{align}
It can be shown that the powers of $\omega$ along the diagonal of $\Ua$ sum to zero modulo $p$. First use
\begin{align*}
\sum_{k=1}^{p-1} \upsilon_k = \frac{p(p-1)}{24}(2 (5 p-1) z^\prime+(p-2) (p^2-1) \gamma^\prime ),
\end{align*}
then note that for primes $p>3$, we have $24|p^2-1$ and $12\vert (p-1)(5p-1)$ so that $\sum_{k=0}^{p-1} \upsilon_k = 0 \bmod p$. Consequently $\det(\Ua)=1$. 

For the $p=3$ case (because of Eq.~\eqref{eqn:DetVF}) we must do a little more work to solve a matrix equation analogous to Eq.~\eqref{eqn:Matrixeqn}. First, we introduce a global phase factor $e^{\i \phi}$ so that $\det \left(e^{\i \phi} \sum_{k=0}^{p-1} \tau^{\gamma k^2} \ketbra{k}{k}\right)=1$ i.e., $\phi=\tfrac{4 \pi \gamma}{9}$. Denote a primitive ninth root of unity as $\zeta=e^{\frac{2 \pi \i}{9}}$ so that
\begin{align*}
\det \left(\zeta^{2\gamma^\prime} C_{\left(\bigl[\begin{smallmatrix} 1 & 0\\ \gamma^\prime & 1 \end{smallmatrix}\bigr]\middle\vert \bigl[\begin{smallmatrix} 1 \\z^\prime\end{smallmatrix}\bigr] \right). }\right)=1
\end{align*}
We must permit our qutrit version of $U_{\pi/8}$ to take a more general form than that given in Eq.~\eqref{eqn:Uadef}, i.e.,
\begin{align*}
U_\upsilon=U(\upsilon_0,\upsilon_1,\ldots)=\sum_{k=0}^{2} \zeta^{\upsilon_k} \ketbra{k}{k} \quad (\upsilon_k \in \mathbb{Z}_9)
\end{align*}
A similar calculation as before leads to the general solution (compare with Eq.~\eqref{eqn:recurrence})
\begin{align}
\upsilon=(0,6z^\prime+2\gamma^\prime+3\epsilon^\prime,6z^\prime+\gamma^\prime+6\epsilon^\prime) \ \bmod{9}\label{eqn:vecdef3}
\end{align}

For example, letting $z^\prime=1,\gamma^\prime=2$ and $\epsilon^\prime=0$
\begin{align}
&\upsilon=(\upsilon_0,\upsilon_1,\upsilon_2)=(0,1,8)\\
&\Rightarrow \Ua(0,1,8)=\left(\begin{array}{ccc}
                                  1 & 0 & 0 \\
                                  0 & \zeta & 0 \\
                                  0 & 0 & \zeta^8  \\
                                \end{array}
                              \right)=\left(\begin{array}{ccc}
                                  1 & 0 & 0 \\
                                  0 & e^{\frac{2 \pi \i}{9}} & 0 \\
                                  0 & 0 & e^{-\frac{2 \pi \i}{9}} \\
                                \end{array}
                              \right)\label{eqn:Uexample3}
\end{align}
One can easily check that all 27 solutions for $z^\prime, \gamma^\prime, \epsilon^\prime \in \Zp$ obey $\sum_{k=0}^{2} \upsilon_k =0 \bmod 3$. However,
\begin{align*}
\det( U_\upsilon)=\zeta^{\sum_{k=0}^{2} \upsilon_k}=\omega^{(z^\prime+\gamma^\prime)}
\end{align*} so that $\zeta^{-(z^\prime+\gamma^\prime)}U_\upsilon \in \SU(3)$.

We finish this section by noting that knowledge of $\Ua D_{(1|0)}\Ua$, $\Ua D_{(0|1)}\Ua$ and Eq.~\eqref{eqn:semidirectproduct} is sufficient to see that the effect (modulo an overall phase) of conjugating an arbitrary Pauli operator by \Ua\ is
\begin{align}
&\Ua D_{(x|z)} \Ua^\dag \propto  C_{\left(\bigl[\begin{smallmatrix} 1 & 0\\ x\gamma^\prime & 1 \end{smallmatrix}\bigr]\middle\vert \bigl[\begin{smallmatrix} x \\ x(z^\prime+2^{-1}\gamma^\prime(x-1))+z \end{smallmatrix}\bigr] \right)}.
\end{align}

%
%

\subsection{Group Structure}\label{sec:Group Structure}

\begin{table}[ht!]
\begin{tabular}{c c |c c c c| c}
  \hline \hline
      & Group        &   \multicolumn{4}{|c|}{No.~elements of order}     &  Min.~no.~of\\
      & name  & $1$  & $p$  & $p^2$  & $p^3$  & generators \\ \hline
$p=2$ & $\mathbb{Z}_8$  & 1 & 1 & 2 & 4 & 1 \\
$p=3$ & $\mathbb{Z}_9 \times \mathbb{Z}_3$ &  1 & 8 & 18 & 0 & 2 \\
$p>3$ & $\mathbb{Z}_p^3$ &  1 & $p^3-1$ & 0 & 0 & 3 \\
\hline
\hline
\end{tabular}
\caption{\label{tab:groupstable} Group structure of the set of diagonal unitaries $\left\{\Ua\right\}$ under matrix multiplication.}
\end{table}
%

For $p>3$ the set $\{\Ua\}\subset \SU(p)$ with matrix multiplication satisfies all the prerequisites to be a group:
\begin{enumerate}
\item Closure: \begin{align*}
&\Ua(z_1,\gamma_1,\epsilon_1)\Ua(z_2,\gamma_2,\epsilon_2)\\&=\Ua(z_1+z_2,\gamma_1+\gamma_2,\epsilon_1+\epsilon_2)
\end{align*}
\item Associativity: Obvious
\item Identity element: $\mathbb{I}=\Ua(0,0,0)$
\item Inverse element: $\Ua^{-1}=U_{-\upsilon}$
\end{enumerate}
The fundamental theorem of finite Abelian groups states that a finite Abelian group is isomorphic to a direct product of cyclic groups of prime-power order. Furthermore, two finite Abelian groups $G,G^\prime$ are isomorphic, $G \cong G^\prime$, if and only if they have identical structure order i.e., if $G$ and $G^\prime$ have the same number of elements of each order. We can use this fact to classify the groups that are generated by all diagonal gates $\{\Ua \vert \forall\ z^\prime, \gamma^\prime, \epsilon^\prime \in \Zp \}$ under matrix multiplication (see Table \ref{tab:groupstable} for a summary). For $p=2$ the gate $U_{\pi/8}$ is sufficient to generate the entire group isomorphic to $\mathbb{Z}_8$. For $p>3$ we have
\begin{align}
(\{\Ua\},\cdot) \cong (\Zp^3,+)
\end{align}
which tells us, amongst other things, that the minimal number of generators required to generate this group is $3$.

For $p=3$ one can check explicitly that the gates $U(z^\prime,\gamma^\prime,\epsilon^\prime)$ form a group, and that
\begin{align}
&\Ua(z_1,\gamma_1,\epsilon_1)\Ua(z_2,\gamma_2,\epsilon_2)=\\
&\begin{cases}
\Ua(z_1+z_2,\gamma_1+\gamma_2,\epsilon_1+\epsilon_2) & \text{if } \gamma_1+\gamma_2<3\\
\Ua(z_1+z_2,\gamma_1+\gamma_2,\epsilon_1+\epsilon_2-1) & \text{if } \gamma_1+\gamma_2\geq 3
\end{cases} \nonumber
\end{align}
The orders of the individual group elements are 
\begin{align}
&\text{ord}(\Ua(0,0,0))=1 \label{eqn:ord1}\\
&\text{ord}(\Ua(z^\prime,0,\epsilon^\prime))=3 \quad (\text{excluding case \eqref{eqn:ord1}}) \label{eqn:ord3} \\
&\text{ord}(\Ua(z^\prime,\gamma^\prime,\epsilon^\prime))=9 \quad (\text{excluding cases \eqref{eqn:ord1} and \eqref{eqn:ord3}}) \nonumber
\end{align}
In fact, for $p=3$ we have
\begin{align}
(\{\Ua\},\cdot) \cong \mathbb{Z}_9 \times \mathbb{Z}_3
\end{align}
i.e., the group is a direct product of cyclic groups of order $9$ and $3$. This group requires only $2$ generators.

The result Eq.~\eqref{eqn:DetVF} has led to an unusual group structure for $p=3$ compared to other primes. Combined with observations from other authors \cite{Rennes:2004,Zhu:2010,Appleby:arxiv09} on SIC-POVMs in $p=3$, perhaps it could be argued that, in the context of quantum information, $3$ is the \emph{second oddest} prime of all \cite{Quotation}.

\section{Geometrical Features}\label{sec:Geometrical Features}


In this section we outline various geometrical properties of the gates $\Ua$ and associated states $\ket{\psi_{\Ua}}$ (as defined below in Eq.~\eqref{eqn:psiU}). Subsection \ref{sec:Useful definitions} provides the various definitions required for this section. In
\ref{sec:Eigenvectors of Clifford Gates} we show that the states $\ket{\psi_{\Ua}}$ are eigenvectors of Clifford gates. This may be of independent interest. In \ref{sec:Noise thresholds for quantum computation} we relate both $\Ua$ and $\ket{\psi_{\Ua}}$ to the convex hulls of Clifford gates and stabilizer states respectively. The reason we do so is twofold: (i) We will see how $\Ua$ and $\ket{\psi_{\Ua}}$ are singled out as being maximally non-Clifford or non-stabilizer in some sense, which is interesting in and of itself (ii) The preceding geometrical property implies that these gates are optimal in the sense of robustness to depolarizing and phase damping noise. This latter property should prove useful in the context of fault-tolerant quantum computation as we discuss in Sec.~\ref{sec:Possible Applications}.

\begin{figure}[ht!]
\begin{center}
\subfigure[]{
\includegraphics[bb=2in 6.5in 5.5in 10.1in, clip=true,scale=0.6]{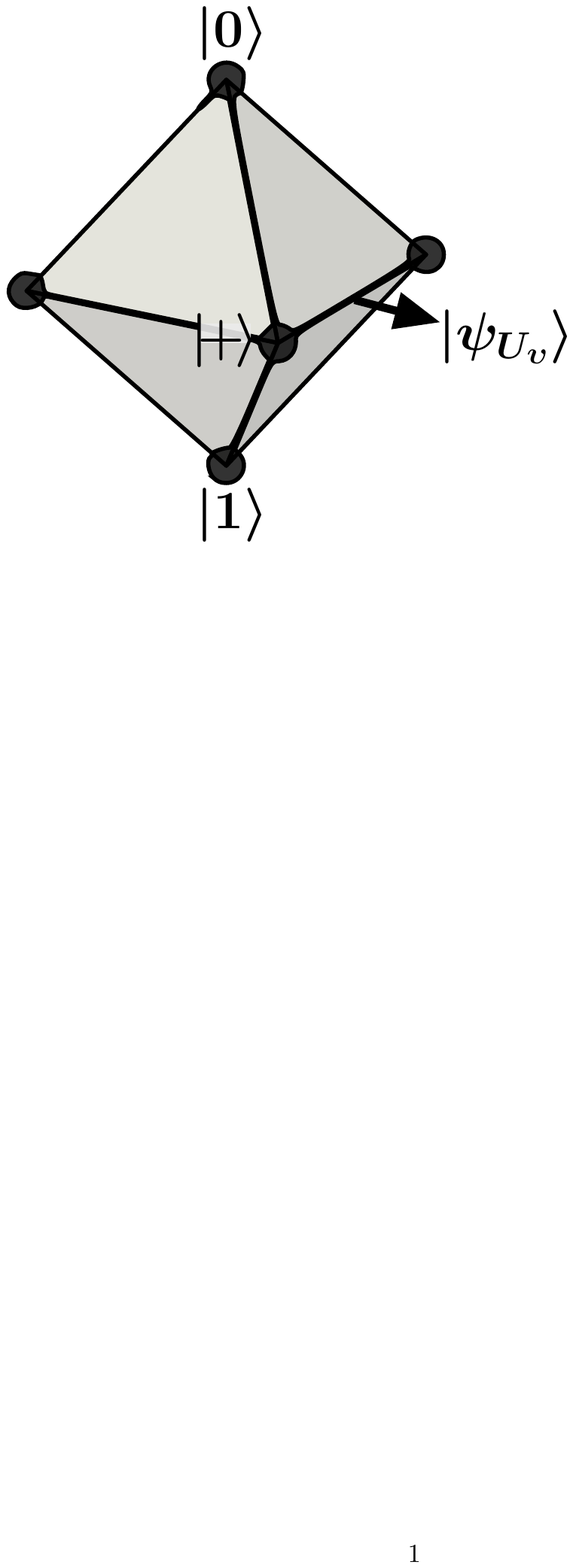}}\\
\subfigure[]{
\includegraphics[bb=3.2in 8.0in 6.9in 10.1in, clip=true,scale=0.5]{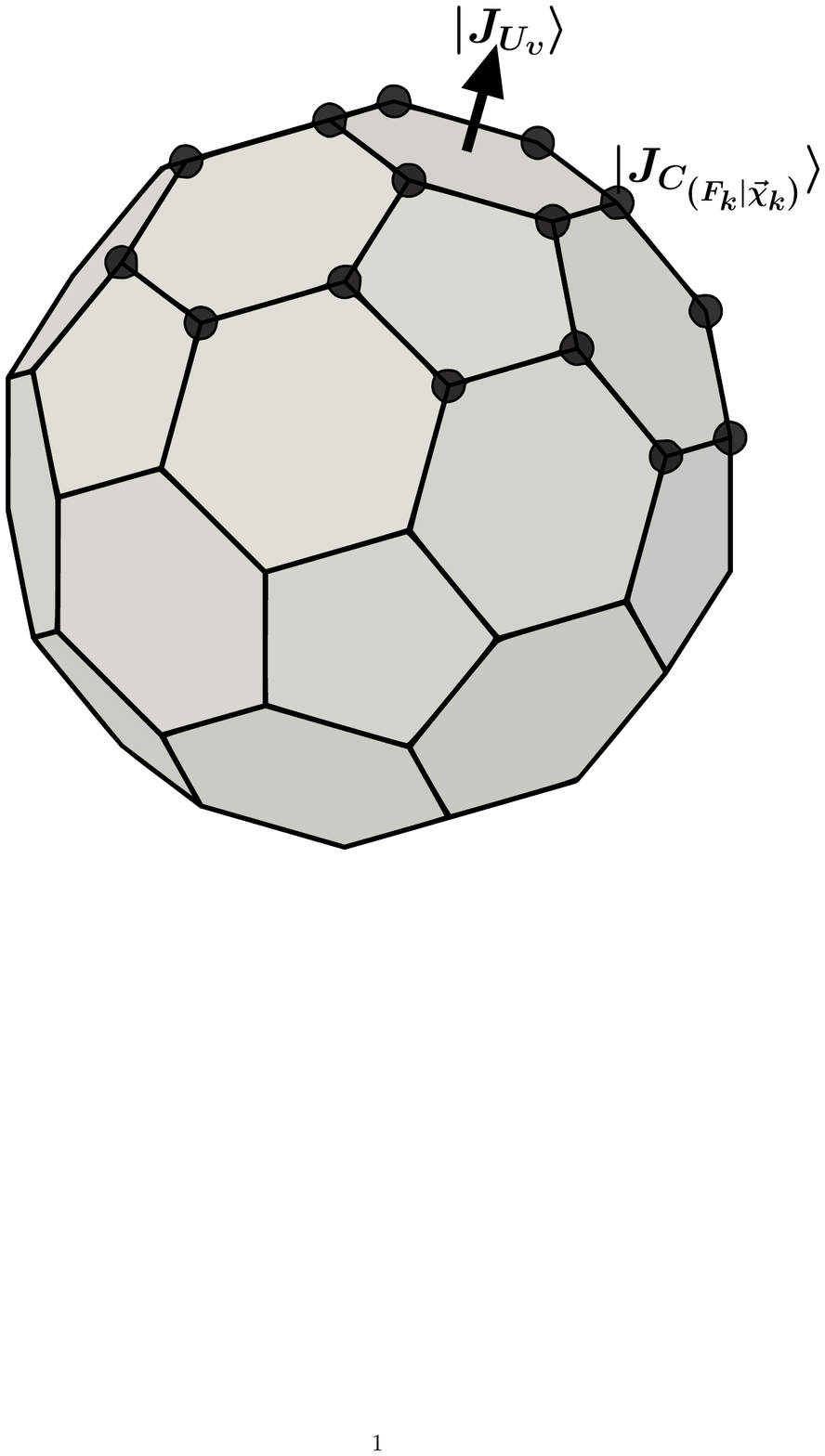}}
\caption{\label{fig:Geometry} Geometry of States and Operations: Filled circles (vertices) correspond to stabilizer states or Clifford gates, respectively. (a) Polytope of Stabilizer States: A single qudit has $p(p+1)$ stabilizer states, the convex hull of which comprises a polytope with $p^{p+1}$ faces. The states $\ket{\psi_{\Ua}}$ are the farthest outside the polytope of all states $\ket{\psi_{U_\theta}}$ (defined in Eq.~\ref{eqn:psiU}). For $p=2$ the stabilizer polytope corresponds to an octahedron (as depicted) with $6$ vertices and $8$ faces. In this dimension $\ket{\psi_{\Ua}}=\ket{H}$ -- the magic state introduced by\cite{Knill05,BravyiKitaev:2005}. Recent results by Campbell \emph{et al.} \cite{Campbell:arxiv12} show that $\ket{\psi_{\Ua}}$ is a magic state for all $p$. (b) Polytope of Clifford Gates: A schematic picture (the object is actually $(p^2-1)^2$-dimensional) of the polytope whose vertices correspond to Clifford gates. It is known with certainty in $p=2$, and with a high degree of plausibility in $p \in \{3,5,7\}$ \cite{WvDMH:2010}, that the gates $\Ua$ are farthest outside the Clifford polytope of all $U \in \U(p)$. Here, each unitary gate, $\Ua$ or $C_{(F\vert \vec{\chi})}\in \mathcal{C}$ (refer to  Eq.~\eqref{eqn:CliffIso}) is represented by its \Jam state (as defined in Eq.~\eqref{eqn:JamIso}).}
\end{center}
\end{figure}

\subsection{Useful definitions}\label{sec:Useful definitions}

As discussed in Sec.~\ref{sec:Group Structure}, the set of gates $\{\Ua\}$ forms a finite group. In fact, they form a finite subgroup, $\{\Ua\}\subset \{U_\theta\}$, of the group of diagonal gates $\{U_\theta\}$ defined as  
\begin{align}
&U_{\theta}=\sum_{k=0}^{p-1} e^{i \theta_k} \ketbra{k}{k} \quad (\theta_k \in \mathbb{R}) \label{eqn:Utheta}\end{align}
We will often have reason to refer to a state $\ket{\psi_{U_\theta}} \in \mathbb{C}^p$ that is very naturally associated with the gate $U_\theta$ via
\begin{align}
&\ket{\psi_{U_{\theta}}}=\frac{1}{\sqrt{p}} \mathrm{diag}(U_{\theta})=U_{\theta}\ket{+} \label{eqn:psiU} \end{align}
where $\ket{+}=(1,1,\ldots,1)/\sqrt{p}$. A \Jam state, $\ket{J_U} \in \mathbb{C}^{p^2}$, corresponding to a unitary operation $U \in \U(p)$ is denoted
\begin{align}
&\ket{J_U}=(\mathbb{I}\otimes U) \sum_{j=0}^{p-1} \frac{\ket{jj}}{\sqrt{p}} \label{eqn:JamIso}
\end{align}
A quantum operation, $\mathcal{E}$, is a superoperator acting upon density operators (i.e., generic quantum states $\rho \in \mathcal{H}_p$) via 
\begin{align}
\mathcal{E}: \rho_{in} \mapsto \rho_{out} \text{ i.e., } \rho_{out}=\mathcal{E}\left(\rho_{in}\right).
\end{align}
The well-known \Jam isomorphism tells us that a complete description of $\mathcal{E}$ is encapsulated in a higher-dimensional state $\varrho_{\mathcal{E}}$ defined as
\begin{align}
\varrho_{\mathcal{E}}=\left[\mathcal{I}\otimes\mathcal{E}\right] \left(\sum_{j,k=0}^{p-1} \frac{\ketbra{j,j}{k,k}}{p}\right)\label{eqn:GenJamIso}
\end{align}
which is the most general form of the operation-state duality given in Eq.~\eqref{eqn:JamIso}.

\subsection{Eigenvectors of Clifford Gates}\label{sec:Eigenvectors of Clifford Gates}

As discussed in Sec.~\ref{sec:Explicitformofquditgates}, we have an explicit form for $p^3$ diagonal gates $\Ua(z^\prime,\gamma^\prime,\epsilon^\prime)$ ( with $z^\prime,\gamma^\prime,\epsilon^\prime \in \Zp$), of which $p^2(p-1)$, corresponding to $\gamma^\prime\neq0$, are non-Clifford. Here we prove that each associated state $\ket{\psi_{U(z^\prime,\gamma^\prime,\epsilon^\prime)}}$ as defined in Eq.~\eqref{eqn:psiU} is an eigenvector of the Clifford operator 
\begin{align*}
C_{\left(\bigl[\begin{smallmatrix} 1 & 0\\ \gamma^\prime & 1 \end{smallmatrix}\bigr]\middle\vert \bigl[\begin{smallmatrix} 1 \\z^\prime\end{smallmatrix}\bigr] \right) }
\end{align*}
with eigenvalue $\omega^{\epsilon^{\prime}}$. Our intuition was that knowledge of such eigenstates should prove useful since both $\ket{T}$ and $\ket{H}$ (qubit magic states as defined in \cite{BravyiKitaev:2005}) are Clifford eigenstates. The recent result by Campbell \emph{et al.} \cite{Campbell:arxiv12} whereby states that we call $\ket{\psi_{\Ua}}$ are shown to be qudit magic 
states, confirms the correctness of this intuition.
In addition, Zauner's conjecture \cite{Zauner:1999} states that fiducial vectors of a Weyl-Heisenberg-covariant SIC-POVM lie in the eigenspace of a particular class of Clifford gates. While the results obtained here are not directly applicable to the resolution of the SIC-POVM problem, they may still prove useful in this context.

To prove the claim, first use Eq.~\eqref{eqn:UDUeqC} to perform the following substitution
\begin{align}
&C_{\left(\bigl[\begin{smallmatrix} 1 & 0\\ \gamma^\prime & 1 \end{smallmatrix}\bigr]\middle\vert \bigl[\begin{smallmatrix} 1 \\z^\prime\end{smallmatrix}\bigr] \right) } \ket{\psi_{U(z^\prime,\gamma^\prime,\epsilon^\prime)}}=\\
&\omega^{-\epsilon} \Ua  D_{(1|0)} \Ua^\dag \ket{\psi_{U(z^\prime,\gamma^\prime,\epsilon^\prime)}}
\end{align}
and subsequently use
\begin{align*}
\Ua^\dag \ket{\psi_{\Ua}}=\Ua^\dag \left(\Ua \ket{+}\right)=\ket{+}
\end{align*}
where $\ket{+}$ is the $+1$ eigenstate of $D_{(1|0)}$, by definition. Clearly, it follows that
\begin{align}
C_{\left(\bigl[\begin{smallmatrix} 1 & 0\\ \gamma^\prime & 1 \end{smallmatrix}\bigr]\middle\vert \bigl[\begin{smallmatrix} 1 \\z^\prime\end{smallmatrix}\bigr] \right) } \ket{\psi_{U(z^\prime,\gamma^\prime,\epsilon^\prime)}}=\omega^{-\epsilon^\prime} \ket{\psi_{U(z^\prime,\gamma^\prime,\epsilon^\prime)}}.
\end{align}

\subsection{Noise thresholds for quantum computation using \Ua \ gates and stabilizer operations}\label{sec:Noise thresholds for quantum computation}

In this section we show how the gates $\Ua$ and states $\ket{\psi_{\Ua}}$ are exceptional with respect to their convex-geometrical relationship to Clifford gates and stabilizer states respectively. We will need to define the stabilizer polytope (the convex hull of stabilizer states), the Clifford polytope (the convex hull of Clifford gates) and a quantity we call negativity (which can be interpreted as a measure of distance outside one of these polytopes). In the present context, a state or gate that is farther outside a polytope generally requires more noise (a higher degree of impurity) to enter said polytope. To this end we introduce a quantity called robustness, which measures the amount of noise that can be tolerated before a gate (state)  becomes expressible as a mixture of Clifford gates (stabilizer states). Table \ref{tab:resultstable} summarizes most of the results in this section. How these results were obtained is explored in the remainder of this subsection.

\begin{table}[ht!]
\begin{tabular}{c c c c c}
  \hline \hline
 & $\varepsilon_D^\star(\Ua)$  & $\varepsilon_{PD}^\star(\Ua)$ & $N(\ket{\psi_{\Ua}})$ & $N(\ket{J_{\Ua}})$ \\\hline
$p=2$ & 45.32\% & 14.65\% & 0.1036 & 2(0.1036)=0.2071\\
$p=3$ & 78.63\% &  36.73\% & 0.1363 & 3(0.1363)=0.4089\\
$p=5$ & 95.24\% &  64.00\% & 0.1600 & 5(0.1600)=0.8000\\
$p=7$ & 97.63\% &  73.27\% & 0.1202 & 7(0.1202)=0.8411\\
\hline
\hline
\end{tabular}
\caption{\label{tab:resultstable}Robustness and negativity: Robustness to noise ($\varepsilon^\star_D$ for depolarizing, $\varepsilon^\star_{PD}$ for phase damping) of a gate $U$ is the noise rate at which a noisy implementation of $U$ enters the Clifford polytope. Negativity can be used as a proxy for distance outside the relevant (stabilizer or Clifford) polytope and is formally defined in Eq.~\eqref{eqn:Negrho} (states) and Eq.~\eqref{eqn:Negvarrho} (gates). A priori, there is no obvious reason why $N(\ket{\psi_{\Ua}})$ and $N(\ket{J_{\Ua}})$ should obey such a simple relationship with one another. In \cite{WvDMH:2010} it was shown that $\Ua$ were the most robust to depolarizing noise of all $U \in \U(p)$ (in dimensions $2$ to $7$ and with some caveats regarding an incomplete facet description of the Clifford polytope). Here we show (for $p\in \{2,3,5,7\}$) that $\Ua$ are also the most robust to phase damping noise of all $U_{\theta}$. The discussion in Sec.~\ref{sec:SimplifiedJamisomorphism} shows that this must also imply that $\ket{\psi_{\Ua}}$ are the most robust states (to depolarizing) of all states $\ket{\psi_{U_\theta}}$. }
\end{table}

\subsubsection{Stabilizer Polytope and Clifford Polytope}

For a single-qudit system there are exactly $p(p+1)$ distinct eigenstates of Pauli operators. For $p=2$ these eigenstates correspond to the vertices of the octahedron depicted in Fig.~\ref{fig:Geometry}(a). The Gottesman-Knill theorem tells us that a supply of stabilizer states, or mixtures of stabilizer states, are useless for the task of promoting a stabilizer circuit to a circuit capable of UQC. As such, the set of all probabilistic mixtures of stabilizer states is an interesting geometric object. It is defined as
\begin{align}
\mathcal{ST\!AB}=&\Bigg\{ \rho\ \Bigg\vert\ \rho=\sum_{i=1}^{p(p+1)} q_i \ketbra{\psi^{(i)}_{_{STAB}}}{\psi^{(i)}_{_{STAB}}}\Bigg\}\nonumber\\
\text{with  } &0\leq q_i \leq 1,\ \sum_{i=1}^{p(p+1)} q_i=1
\end{align}
where $q_i$ can be understood as probabilities and $\ket{\psi^{(i)}_{_{STAB}}}$ are the aforementioned Pauli eigenstates (stabilizer states).

The convex hull of a finite set of points forms what is generally known as a polytope (a higher-dimensional generalization of a polyhedron). Using the Minkowski-Weyl theorem, we know that every such polytope has a description in terms of a finite number of facets (bounding inequalities, also known as halfspaces). In the present context, it is known \cite{Cormick:2006} that exactly $p^{p+1}$ facets (which we denote $A$) are required to describe $\mathcal{ST\!AB}$. More precisely, testing a state $\rho$ for membership of the polytope $\mathcal{ST\!AB}$, leads to the following condition
\begin{align}
\rho \in \mathcal{ST\!AB} \iff \underset{u \in \Zp^{p+1}}{\min}  \Tr\left[A(u) \rho \right] \geq 0.
\end{align}
If a state $\rho$ is outside $\mathcal{ST\!AB}$ then we define the negativity \cite{Negativity} of the state, $N(\rho)$ as
\begin{align}
N(\rho)=\lvert \underset{u \in \Zp^{p+1} }{\min} \Tr\left[A(u) \rho \right] \rvert .\label{eqn:Negrho}
\end{align}
The threshold depolarizing rate, $\varepsilon_D^\star(\ket{\psi_{U_{\theta}}})$, of a state, $\ket{\psi_{U_{\theta}}}$, is the minimum value of $\varepsilon_D$ required to make $\ket{\psi_{U_{\theta}}}$ an element of $\mathcal{ST\!AB}$:
\begin{align}
&\varepsilon_D^\star(\ket{\psi_{U_{\theta}}})=\min \varepsilon_D\ (0 \leq \varepsilon_D \leq 1) \text{ such that}\\
&(1-\varepsilon_D) \ketbra{\psi_{U_{\theta}}}{\psi_{U_{\theta}}} + \varepsilon_D \tfrac{\mathbb{I}}{p} \in \mathcal{ST\!AB}.\nonumber
\end{align}
The quantities $N(\rho)$ and $\varepsilon_D^\star(\rho)$ are (inversely) related, as discussed in \cite{WvDMH:2010}.

An explicit definition for individual facets $A(u)$ is given by
\begin{align}
A({u})=\frac{1}{p}\left( \Pi_{(0|1)[u_0]}+\sum_{j=1}^{p}\Pi_{(1|j-1)[u_j]} -\I \right) \label{eqn:Aproj}
\end{align}
where  $\Pi_{(a|b)[k]}$ is the projector onto the $\omega^k$ eigenspace of $X^aZ^b$ i.e.,
\begin{align}
\Pi_{(a|b)[k]}=\frac{1}{d}\left(I+\w^{-k}X^aZ^b+\ldots +\w^{-(p-1)k}(X^aZ^b)^{p-1}\right) \label{Qproj}
\end{align}

Using the \Jam isomorphisms of Eq.~\eqref{eqn:JamIso} and Eq.~\eqref{eqn:GenJamIso} we can construct an object (polytope) that is analogous to $\mathcal{ST\!AB}$, but where the vertices now correspond to Clifford gates rather than stabilizer states. As before, quantum operations that are expressible as a mixture of Clifford operations are useless for the task of promoting a stabilizer circuit to a circuit that is capable of UQC. We denote this so-called Clifford polytope as $\mathcal{CLIFF}$ \cite{Buhrmanetal:2006,WvDMH:2010}

\begin{align}
&\mathcal{CLIFF}=\Bigg\{ \varrho_{\mathcal{E}}\ \Bigg\vert\ \nonumber \\ &\qquad \qquad \varrho_{\mathcal{E}}=\sum_{j=1,k=1}^{j=p(p^2-1),k=p^2} q_{j,k} \ketbra{J_{C_{\left(F_j\vert \vec{\chi}_k\right)}}}{J_{C_{\left(F_j\vert \vec{\chi}_k\right)}}}\Bigg\}\nonumber\\
&\qquad\text{with  }\quad 0\leq q_{j,k} \leq 1,\ \sum_{j=1,k=1}^{j=p(p^2-1),k=p^2} q_{j,k}=1
\end{align}
Testing an arbitrary quantum operation $\mathcal{E}$ for membership of the Clifford polytope requires construction of the associated \Jam state $\varrho_{\mathcal{E}}$ (as described in Eq.~\eqref{eqn:GenJamIso}) and then using 
\begin{align}
\varrho_{\mathcal{E}} \in \mathcal{CLIFF} \iff \Tr\left(W \varrho_{\mathcal{E}}\right) \geq 0\quad (\forall W \in \mathcal{W}),
\end{align}
where $\mathcal{W}$ is a finite set of facets describing $\mathcal{CLIFF}$.  Analogously to Eq.~\eqref{eqn:Negrho}, we define the negativity of an operation $\mathcal{E}$ as
\begin{align}
N(\varrho_{\mathcal{E}})=\lvert \underset{W \in \mathcal{W} }{\min} \Tr\left[W \varrho_{\mathcal{E}} \right] \rvert .\label{eqn:Negvarrho}
\end{align}
The threshold depolarizing rate, $\varepsilon_D^\star(U_{\theta})$, of a gate, $U_{\theta}$, is the minimum value of $\varepsilon_D$ required to make $U_{\theta}$ an element of $\mathcal{CLIFF}$:
\begin{align}
&\varepsilon_D^\star(U_{\theta})=\min \varepsilon_D\ (0 \leq \varepsilon_D \leq 1) \text{ such that} \label{eqn:robustnessgate}\\
&(1-\varepsilon_D) \ketbra{J_{U_{\theta}}}{J_{U_{\theta}}} + \varepsilon_D \tfrac{\mathbb{I}}{p^2} \in \mathcal{CLIFF}.\nonumber
\end{align}
While $\mathcal{W}$ is known to exist, and be finite, the complexity of halfspace enumeration is such that we can only claim to have derived in \cite{WvDMH:2010} (at least) a subset of $\mathcal{W}$. Nevertheless, if a given $\varrho_{\mathcal{E}}$ (encoding an operation $\mathcal{E}$) satisfies $\Tr\left(W \varrho_{\mathcal{E}}\right) < 0$ for some $W \in \mathcal{W}$, then this operation is unambiguously outside the Clifford polytope.

\subsubsection{Robustness to Depolarizing Noise i.e.,\\ Maximally Non-Clifford Gates}\label{sec:RobustnesstoDepolarizing Noise}

In \cite{WvDMH:2010} a gate $U_{opt}\in \U(p)$ was found, for each of $p \in \{2,3,5,7\}$, which required very high amounts of depolarizing noise to become expressible as a mixture of Clifford gates. There, it was suggested that the simple form of $U_{opt}$ and their high robustness to noise (i.e., high $\varepsilon^\star_D$ in Eq.~\eqref{eqn:robustnessgate}) made them analogous to the qubit $U_{\pi/8}$ gate in some sense. Here we strengthen the analogy by showing that $U_{opt}$ are actually equivalent (i.e., the same up to a factor of a Clifford gate) to the gates $\Ua$ that we have derived by enforcing that they should be diagonal elements of $\mathcal{C}_3$.

The state $\ket{J_{U_{opt}}}$ that is farthest outside $\mathcal{CLIFF}$ (i.e. the convex polytope whose vertices are Clifford gates) is that state which achieves
\begin{align}
\underset{W \in \mathcal{W}, U \in \SU(p)}{\min}\Tr\left( W \ketbra{J_U}{J_U} \right) 
\end{align}
where $\mathcal{W}$ is the bounding set of facets that describes $\mathcal{CLIFF}$. Here we give the explicit relationship between highly (and maybe maximally) robust gates given in \cite{WvDMH:2010} and the generalized versions of $U_{\pi/8}$ that we have described in Sec.~\ref{sec:Explicitformofquditgates}:
\begin{align*}
&p=2: \qquad U_{opt}=U_{\pi/8}\\
&p=3: \qquad U_{opt}=\Ua C_{\left(\bigl[\begin{smallmatrix} -1 & 0\\ 0 & -1 \end{smallmatrix}\bigr]\middle\vert \bigl[\begin{smallmatrix} 0 \\ 0 \end{smallmatrix}\bigr] \right) } \left(\Ua\text{ in Eq.~}\eqref{eqn:Uexample3}\right)\\
&p=5: \qquad U_{opt}=\Ua C_{\left(\bigl[\begin{smallmatrix} -1 & 0\\ -1 & -1 \end{smallmatrix}\bigr]\middle\vert \bigl[\begin{smallmatrix} 0 \\ 3 \end{smallmatrix}\bigr] \right) } \left(\Ua\text{ in Eq.~}\eqref{eqn:Uexample5}\right)\\
&p=7: \qquad U_{opt}=\Ua C_{\left(\bigl[\begin{smallmatrix} -1 & 0\\ 2 & -1 \end{smallmatrix}\bigr]\middle\vert \bigl[\begin{smallmatrix} 0 \\ 2 \end{smallmatrix}\bigr] \right) }\\
&\qquad \qquad  \qquad [\text{ with }\Ua=\Ua(z^\prime=1,\gamma^\prime=2,\epsilon^\prime=0)]
\end{align*}
The Clifford gates that relate $U_{opt}$ and $\Ua$ can be absorbed into $W$, creating another $W^\prime \in \mathcal{W}$ with the same spectrum, and
\begin{align}
\Tr\left( W^\prime \ketbra{J_{\Ua}}{J_{\Ua}} \right) = \Tr\left( W \ketbra{J_{\Ua C}}{J_{\Ua C}} \right) 
\end{align}
so that the negativity and robustness results from  \cite{WvDMH:2010} apply here too.

\subsubsection{Phase Damping Thresholds via Simplified \Jam Isomorphism}\label{sec:SimplifiedJamisomorphism}

Phase damping is a physically well-motivated noise process (often interpreted as resulting from so-called phase kicks), whose overall effect on a state $\rho$ is to uniformly decrease the amplitude of all off-diagonal elements in $\rho$ (see e.g.,  \cite{Byrd:2011}). The implementation of a diagonal gate $U_\theta$, while suffering phase damping noise (with noise rate $\varepsilon_{PD}$), results in an overall operation
\begin{align}
\mathcal{E}(\rho)&=(1-\varepsilon_{PD})U_\theta \rho U_\theta^\dag + \frac{\varepsilon_{PD}}{p-1} \sum_{k=1}^{p-1} \left(Z^k U_\theta\right) \rho \left(Z^k U_\theta\right)^\dag \label{eqn:PDdefn}\\
&=(1-\varepsilon_{PD})U_\theta \rho U_\theta^\dag + \frac{\varepsilon_{PD}}{p-1} \left(\mathbb{I}-U_\theta \rho U_\theta^\dag\right)
\end{align}
so that the robustness to phase damping noise, $\varepsilon^\star_{PD}(U_\theta)$, of a gate, $U_\theta$, is 
\begin{align}
&\varepsilon_{PD}^\star(U_{\theta})=\min \varepsilon_{PD}\ (0 \leq \varepsilon_{PD} \leq 1) \text{ such that} \label{eqn:robustnessgate}\\
&\varrho_\mathcal{E} \in \mathcal{CLIFF} \qquad  \mathcal{E}\text{ as in Eq.~\eqref{eqn:PDdefn}} \nonumber
\end{align}

Given a diagonal gate $U_{\theta}$ (as defined in Eq.~\eqref{eqn:Utheta}) it is trivial to see that the state $\ket{\psi_{U_\theta}}$ (as defined in Eq.~\eqref{eqn:psiU}) provides a complete description of the gate. A quantum process consisting of a probabilistic mixture of various different diagonal gates $U_{\theta}$ is thus representable by a quantum state
\begin{align}
\rho=\sum_{i} q_i \ketbra{\psi_{U_\theta}^{(i)}}{\psi_{U_\theta}^{(i)}} \quad \left(0\leq q_i \leq 1, \sum_i q_i=1\right) \label{eqn:Equatorialstates}
\end{align}
Effectively, this is a simplified form of the \Jam isomorphism (given in Eq.~\eqref{eqn:GenJamIso}) that is only possible because our allowed operations are highly restricted. With that said, the operation in Eq.~\eqref{eqn:PDdefn} can equally well be represented by a single-qudit state:
\begin{align*}
\rho &=(1-\varepsilon_{PD})\ketbra{\psi_{U_\theta}}{\psi_{U_\theta}}+\frac{\varepsilon_{PD}}{p-1}\sum_{k=1}^{p-1} \ketbra{\psi_{Z^k U_\theta}}{\psi_{Z^k U_\theta}}\\
&=(1-\varepsilon_{PD})\ketbra{\psi_{U_\theta}}{\psi_{U_\theta}}+\frac{\varepsilon_{PD}}{p-1}\left(\mathbb{I}-\ketbra{\psi_{U_\theta}}{\psi_{U_\theta}}\right)
\end{align*}
To summarize: a noisy (phase-damped) implementation of $U_\theta$ can be identified as a state of the form Eq.~\eqref{eqn:Equatorialstates}.

As discussed in Sec.~\ref{sec:Explicitformofquditgates} there are exactly $p^2$ diagonal Clifford gates, corresponding to $\Ua(z^\prime,0,\epsilon^\prime)$, and each such gate corresponds to a state $\ket{\psi_{\Ua}}$. These $p^2$ states comprise the vertices of a polytope of dimension $p(p-1)$ contained in the space spanned by Eq.~\eqref{eqn:Equatorialstates}. For example (and consulting Fig.~\ref{fig:Geometry}(a)), in the $p=2$ case one sees that the (two-dimensional) \protect{$x$-$y$} plane of the octahedron contains $p^2=4$ vertices. Checking the negativity (distance outside this polytope) of a given state amounts to evaluating its expectation value with respect to all $p^p=4$ facets that comprise the polytope boundary (i.e., the four edges of the octahedron that are contained in the $x$-$y$ plane). Of all states $\ket{\psi_{U_\theta}}$, the one farthest outside this polytope is $\ket{\psi_{U_{\pi/8}}}$. The general expression for all $p^p$ distinct facets of this $p(p-1)$-dimensional polytope (which we call $A_{\text{edge}}$) is
\begin{align}
A_{\text{edge}}(u_1,u_2,\ldots,u_p)&=\frac{1}{p}\sum_{u_0=0}^{p-1}A(u_0,u_1,\ldots,u_p)\\
&=\frac{1}{p}\left(\sum_{j=1}^{p}\Pi_{(1|j-1)[u_j]} -\frac{\I}{p} \right)
\end{align}
where $A$ are the facets of $\mathcal{ST\!AB}$ as defined in Eq.~\eqref{eqn:Aproj}.

The final element that is required is to realize that an operation $U_\theta$ that is maximally robust to phase damping noise is exactly the operation for which $\ket{\psi_{U_\theta}}$ is most resistant to depolarizing noise before entering the $\protect{p(p-1)}$-dimensional polytope discussed above. In fact, a simple calculation shows that
\begin{align*}
\varepsilon^\star_{PD}(U_\theta) =\frac{p-1}{p} \varepsilon^\star_{D}(\ket{\psi_{U_\theta}})
\end{align*}
where $\varepsilon^\star_{PD}(U_\theta)$ is the phase-damping noise rate required to make $U_\theta$ enter the convex hull of diagonal Clifford gates, and $\varepsilon^\star_{D}(\ket{\psi_{U_\theta}})$ is the depolarizing noise rate required to make $\ket{\psi_{U_\theta}}$ enter the convex hull of stabilizer states.

Any facet $A_{\text{edge}}$ can be decomposed as
\begin{align}
A_{\text{edge}}=\sum_j \lambda_j \ketbra{\lambda_j}{\lambda_j} \quad (\lambda_1 \leq \lambda_2 \ldots ),
\end{align}
and the state $\ket{\lambda_1}$ is the state that, out of all qudit states $\rho$, maximally violates $\Tr\left[A_{\text{edge}} \rho\right]\geq 0$. A simple calculation shows that the minimum eigenvalue of $A_{\text{edge}}$ (over all possible $A_{\text{edge}}$) is $\lambda_{1}(A_{\text{edge}})=-(p-1)/p^2$, when $p$ is an odd prime. Consequently, if $\lambda_{1}=-(p-1)/p^2$, and $\ket{\lambda_{1}}$ is of the form $\ket{\psi_{U_\theta}}$, then $\ket{\lambda_{1}}$ is maximally robust to depolarizing noise and the operation $U_\theta$ that $\ket{\lambda_{1}}$ represents is maximally robust to phase-damping noise. This is the case for $p \in \{2,5\}$ in our current investigation. For $p \in \{3,7\}$, the states $\ket{\lambda_{1}}$ that achieve $\lambda_{1}=-(p-1)/p^2$ are not of the form $\ket{\psi_{U_\theta}}$ and so we had to resort to a numerical optimization over all $A_{\text{edge}}$ and all states $\ket{\psi_{U_\theta}}$.

For $p=3$ there are two distinct types of edge, as classified by spectrum:
\begin{align*}
&\lambda(A_\text{edge})=\left\{-\tfrac{2}{9},\tfrac{1}{9},\tfrac{4}{9}\right\}\\
\text{or} \\
&\lambda(A_\text{edge})=\Big\{\frac{1}{9}\left(3\sin\tfrac{\pi}{18}-\sqrt{3}\cos\tfrac{\pi}{18}\right),\\&\quad  \frac{1}{9}\left(1+3\sin\tfrac{\pi}{18}-\sqrt{3}\cos\tfrac{\pi}{18}\right),\\&\quad  \frac{1}{9}\left(1+2\sqrt{3}\cos\tfrac{\pi}{18}\right)\Big\}
\end{align*}
There are $p^2(p-1)=18$ of the latter $A_\text{edge}$, where the minimizing eigenvector for each distinct facet corresponds to a distinct non-Clifford $\Ua$. In $p=5$ there are exactly $p^2(p-1)=100$ edges with spectrum
\begin{align*}
\lambda(A_\text{edge})=\{-0.16,-0.08361,0.04,0.04,0.36361\}
\end{align*}
and these correspond to the $100$ non-Clifford $\Ua$. In $p=7$ there are at least $2(7^2)=98$ facets $A_\text{edge}$ with minimal eigenvalue $-0.12016$, whose corresponding eigenvector is of the form $\ket{\psi_{\Ua}}$. These were the most robust states of all $\ket{\psi_{U_\theta}}$ that we could find by optimization, but it is possible that we became trapped in a local minimum.

As a final comment, we note that qubit form of this argument (simplified \Jam isomorphism etc.) was presented by Virmani \emph{et al.} \cite{VirmaniHuelgaPlenio:2005} (see also \cite{Plenio2010}) where an adversarial phase damping model was used to obtain upper bounds on the quantum fault-tolerance threshold. In that case it was found that 
\begin{align}
\left[14.7\% \approx \varepsilon^\star_{PD}(U_{\pi/8})\right] =\frac{1}{2}\left[ \varepsilon^\star_{D}(\ket{\psi_{U_{\pi/8}}})\approx 29.3\%\right],
\end{align}
which implies that the $U_{\pi/8}$ gate, whilst maximally robust amongst all diagonal gates, requires about $15\%$ phase damping noise before it becomes expressible as a mixture of (diagonal) Clifford gates.

\section{Applications in Fault-Tolerant Quantum Computing}\label{sec:Possible Applications}

In \cite{Campbell:arxiv12} Campbell \emph{et al.} quote results by Nebe, Rains and Sloane \cite{NRS:2001,NRS:2006} which show that the gate set
\begin{align}
\left\langle \mathcal{C}, \textsc{CSUM}, U \right\rangle
\end{align}
is dense in $\SU(p^n)$, where $\mathcal{C}$ is the set of single-qudit Cliffords, \textsc{CSUM} is the generalized version of the \textsc{CNOT} gate and $U$ is a non-Clifford single-qudit gate,
\begin{align}
&\textsc{CSUM}:\ket{a}\ket{b} \mapsto \ket{a}\ket{a+b \bmod p}\\
&U \in \SU(p)\backslash \mathcal{C}.
\end{align}
In particular, $\Ua$ is sufficient to promote multi-qudit Clifford gates to a universal gate set. We expect that $\Ua$ possesses all the additional qualities that makes $U_{\pi/8}$ the preferred non-Clifford gate in qubit-based universal gate sets.

In \cite{Gottesman1999a,Boykin2000} it is argued, for the qubit case, that creation of a Clifford eigenstate should be easier to do in a fault-tolerant manner than fault-tolerant implementation of $\Ua$ directly. It is easy to see that, for an arbitrary qudit state $\ket{\psi_{\text{arb}}}$,
\begin{align*}
\Pi_{(0,0|1,d-1)[0]} (\ket{\psi_{\Ua}} \otimes \ket{\psi_{\text{arb}}})=(\Ua\ket{\psi_{\text{arb}}})\otimes\ket{0} \label{eqn:gateinject}
\end{align*}
where $\Pi_{(0,0|1,d-1)[0]}$ denotes a rank-$p$ projector onto the $\omega^0$ eigenspace of the operator $Z\otimes Z^{-1}$ (i.e., a stabilizer measurement). Clearly, creation of Clifford eigenstates $\ket{\psi_{\Ua}}$ (see Sec.~\ref{sec:Eigenvectors of Clifford Gates}) is sufficient to promote a stabilizer circuit to UQC in the qudit case too.

In \cite{Campbell:arxiv12} Campbell \emph{et al.} introduce a qudit gate $M$ of the form
\begin{align}
&M^{(p)}=\sum_{j=0}^{p-1} e^{\frac{2 \pi i}{p^2} \lambda_j} \ketbra{j}{j}\\
\text{with } &\lambda_j=p{j \choose 3}-j{p \choose 3}+{p+1 \choose 4}
\end{align}
Note that 
\begin{align}
&M^{(3)}=\Ua(z^\prime=1,\gamma^\prime=1,\epsilon^\prime=0)\\
&M^{(5)}=\Ua(z^\prime=2,\gamma^\prime=1,\epsilon^\prime=2)\\
&M^{(7)}=\Ua(z^\prime=3,\gamma^\prime=1,\epsilon^\prime=4)
\end{align}
in our notation. They showed that $\ket{M^{(p)}_0}$, which correspond to $\ket{\psi_{\Ua}}$ defined in Eq.~\eqref{eqn:psiU}, are distillable for all prime dimensions and hence are magic states. For $p=3$ a distillation routine with remarkably good performance was found and we use this result in the discussion below.

\begin{figure*}[ht!]
\begin{center}
\subfigure[]{
\includegraphics[scale=1.2]{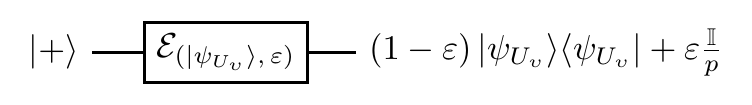}}\\
\subfigure[]{
\includegraphics[scale=1.2]{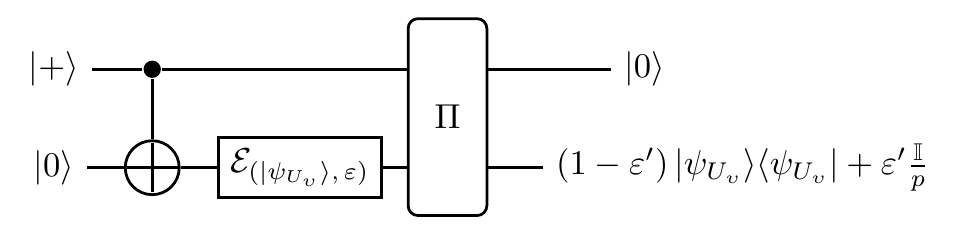}}
\caption{\label{fig:Dilution} Dilution of Noise in Magic State Preparation: (a)Straightforward Magic State Preparation: Using the superoperator $\mathcal{E}_{(\ket{\psi_{\Ua}},\varepsilon)}$ to prepare an imperfect (depolarized) version of a magic state $\ket{\psi_{\Ua}}$. (b)Postselected Magic State Preparation: Using the same superoperator $\mathcal{E}_{(\ket{\psi_{\Ua}},\varepsilon)}$ to create a less imperfect ($\varepsilon^\prime < \varepsilon$) version of the magic state $\ket{\psi_{\Ua}}$. The Pauli measurement operator $\Pi$ stands for postselection on receiving the outcome $+1$ using  the measurement $\Pi_{(0,0|1,p-1)[0] }$ (see Eq.~\eqref{eqn:gateinject}). The circuit elements to the left of $\Pi$ implement the creation of the \Jam state $\varrho_{\mathcal{E}}$ (see Eq.~\eqref{eqn:GenJamIso}) describing $\mathcal{E}_{(\ket{\psi_{\Ua}},\varepsilon)}$. }
\end{center}
\end{figure*}

If we have access to a superoperator $\mathcal{E}_{(\ket{\psi_{\Ua}},\varepsilon)}$ defined as 
\begin{align}
\mathcal{E}_{(\ket{\psi_{\Ua}},\varepsilon)}(\rho)=(1-\varepsilon)\Ua \rho \Ua^\dag + \varepsilon \frac{\mathbb{I}}{p}\label{eqn:depolU}
\end{align}
then it can be used to create noisy versions of $\ket{\psi_{\Ua}}$ in a straightforward way, as depicted in Fig.~\ref{fig:Dilution}(a). However, a simple circuit given in Fig.~\ref{fig:Dilution}(b) shows how a less noisy version of $\ket{\psi_{\Ua}}$ can be created by using the same operation $\mathcal{E}_{(\ket{\psi_{\Ua}},\varepsilon)}$ as well as some additional stabilizer operations. The straightforward method produces a state $\ket{\psi_{\Ua}}$ with effective depolarization rate $\varepsilon$ whereas the postselected version produces $\ket{\psi_{\Ua}}$ with effective depolarization rate $\varepsilon^\prime (<\varepsilon)$. The relationship between the effective noise rates is
\begin{align}
\varepsilon=\frac{p \varepsilon^\prime}{1+(p-1)\varepsilon^\prime}
\end{align}
and, inverted,
\begin{align}
\varepsilon^\prime=\frac{\varepsilon}{p-(p-1)\varepsilon}.\label{eqn:dilution}
\end{align}

\begin{table}[ht!]
\begin{tabular}{c c c}
  \hline \hline
 & Lower Bound  & Upper Bound  \\\hline
$p=2$ & 45.32\% & 45.32\%\\
$p=3$ & 58.15\% &  78.63\% \\
$p=5$ & 80.61\% &  95.20\% \\
$p=7$ & 72.24\% &  97.63\% \\
\hline
\hline
\end{tabular}
\caption{\label{tab:dilutiontable} Noise thresholds for universal quantum computation: When supplementing the set of Clifford gates, which are presumed to be perfect, with a depolarized version of $\Ua$ (as in Eq.~\eqref{eqn:depolU}), then there exist regimes of the noise parameter, $\varepsilon$, for which UQC is either provably possible or provable impossible. Noise rates between these two bounds are those for which we currently have no proof regarding the utility of depolarized $\Ua$  when supplementing stabilizer operations. Lower bounds are evaluated by using the results of Campbell \emph{et al.}\cite{Campbell:arxiv12}, along with the noise dilution protocol of Fig.~\ref{fig:Dilution} and Eq.~\eqref{eqn:dilution}. Upper bounds are established via geometrical arguments in Sec.~\ref{sec:RobustnesstoDepolarizing Noise} and \cite{WvDMH:2010}. }
\end{table}

In \cite{Campbell:arxiv12} it was shown that, for $p=3$, depolarized versions of $\ket{\psi_{\Ua}}$ could be distilled for noise rates up to about $32\%$. The implication is that a superoperator $\mathcal{E}_{(\ket{\psi_{\Ua}},\varepsilon)}$ enables universal quantum computation (via MSD) up to noise rates of around $58\%$. This is found by solving
\begin{align}
\frac{\varepsilon}{3-2\varepsilon}&=0.3165\\
\Rightarrow \qquad \varepsilon &= 0.5815
\end{align}
Our results in Sec.~\ref{sec:RobustnesstoDepolarizing Noise} and \cite{WvDMH:2010} indicate that $\mathcal{E}_{(\ket{\psi_{\Ua}},\varepsilon)}$ can never enable universal quantum computation (by supplementing Clifford gates) for noise rates $\epsilon \geq 78.6\%$. In the qubit case it was shown, using similar arguments, by Reichardt\cite{ReichardtMagic09} and Buhrman \emph{et al.} \cite{Buhrmanetal:2006} (and more generally in \cite{WvDMH:2009}), that the noise rates for which $\mathcal{E}_{(\ket{\psi_{U_{\pi/8}}},\varepsilon)}$ was outside the Clifford polytope were exactly those noise rates for which $\mathcal{E}_{(\ket{\psi_{U_{\pi/8}}},\varepsilon)}$ could supplement Clifford gates to enable universality. It is an interesting open question  for qutrit systems whether the operation $\mathcal{E}_{(\ket{\psi_{\Ua}},\varepsilon)}$ with noise rates in the range $\protect{58.15\% < \varepsilon < 78.6\%}$ can enable universal quantum computation. All of the above techniques can, in principle, be applied to any prime dimensional $\ket{\psi_{\Ua}}$ and we present a summary in Table.~\ref{tab:dilutiontable} for dimensions $p \in\{2,3,5,7\}$. Indeed, the performance of the $p=5$ MSD routine is such that the gap between upper and lower bounds in Table.~\ref{tab:dilutiontable} is even smaller in this case. If additional qudit MSD routines are developed, then the lower bounds of Table.~\ref{tab:dilutiontable} can potentially be raised. Similarly, if an analagous scenario to that of bound non-stabilizer states \cite{CampbellBrowne:2010,Veitch:arxiv12} also holds for operations, then the upper bounds in Table.~\ref{tab:dilutiontable} can potentially be lowered.

\section{Summary \& Open Questions}
Motivated by the utility and geometric prominence of the qubit $U_{\pi/8}$ gate, we provided an explicit solution for all diagonal qudit gates that displayed the same relationship with the Clifford group (i.e., we constructed diagonal gates from the third level of the Clifford hierarchy). We saw that these diagonal gates generated a finite group whose structure depended upon whether $p=2$, $p=3$ or $p>3$. It might be interesting to fully enumerate all the single-qudit elements of $\mathcal{C}_3$, or analyze the structure of the diagonal subset of $\mathcal{C}_k$ (particularly for $p=3$). Geometrically, these generalized $U_{\pi/8}$ gates, which we have called $\Ua$, appear to display the same relationship with the set of Clifford gates -- a relationship which makes them maximally non-Clifford in some sense. The state $\ket{\psi_{\Ua}} \in \mathbb{C}^p$, defined as $\ket{\psi_{\Ua}}=\Ua\ket{+}$, was already known to be useful and geometrically significant in the $p=2$ case (where it is widely known as the $\ket{H}$-type magic state), and we discussed some properties of the general qudit case which led us to believe they could also be useful. As we completed this work we became aware of results by Campbell \emph{et al.} \cite{Campbell:arxiv12} which show that 
states $\ket{\psi_{\Ua}}$ are indeed magic states for all prime dimensions. In the final section we use the results of Campbell \emph{et al.} to show noise rates for which noisy versions of $\Ua$ can and cannot provide UQC (when supplementing the full set of Clifford gates). A very interesting problem is to further close the gap between these noise regimes, a gap that is non-existent in the qubit case.

\section{Acknowledgements}
We thank E.~Campbell for helpful comments on a previous
version of this manuscript. M.H. was supported by the Irish
Research Council as an Empower Fellow. J.V.~acknowledges support from Science Foundation Ireland under the Principal Investigator Award 10/IN.1/I3013


\begin{thebibliography}{99}

\bibitem{Gottesman:1998}
D.~Gottesman,
``Theory of fault-tolerant quantum computation''
\newblock Phys.~Rev.~A \textbf{57}, 127 (1998).
\href{http://dx.doi.org/10.1103/PhysRevA.57.127}{http://dx.doi.org/10.1103/PhysRevA.57.127}

\bibitem{Boykin2000}
P.~O.~Boykin, T.~Mor, M.~Pulver, V.~Roychowdhury and F.~Vatan.
``A new universal and fault-tolerant quantum basis''
\newblock Information Processing Letters \textbf{75}, 3 pp.~101--107, (2000).
\href{http://dx.doi.org/10.1016/S0020-0190(00)00084-3}{http://dx.doi.org/10.1016/S0020-0190(00)00084-3}

\bibitem{Buhrman2001}
H.~Buhrman, R.~Cleve and W.~van Dam,
``Quantum entanglement and communication complexity''
\newblock SIAM \textbf{30}, 6 pp.~1829--1841, (2001).
\href{http://dx.doi.org/10.1137/S0097539797324886}{http://dx.doi.org/10.1137/S0097539797324886}
\bibitem{Howard2012}
M.~Howard and J.~Vala,
``Nonlocality as a benchmark for universal quantum computation in Ising anyon topological quantum computers'',
\newblock Phys.~Rev.~A \textbf{85}, 022304 (2012).
\href{http://dx.doi.org/10.1103/PhysRevA.85.022304}{http://dx.doi.org/10.1103/PhysRevA.85.022304}


\bibitem{Broadbent:2009}
A.~Broadbent, J.~Fitzsimons  and E.~Kashefi,
``Universal blind quantum computation'',
 \emph{Annual IEEE Symposium on Foundations of Computer Science,}
 pp.~517 --526, (2009).
\href{http://dx.doi.org/10.1109/FOCS.2009.36}{http://dx.doi.org/10.1109/FOCS.2009.36} 


\bibitem{Gottesman1999b}
D.~Gottesman and I.~L.~Chuang,
``Demonstrating the viability of universal quantum computation using teleportation and single-qubit operations''
\newblock Nature \textbf{402}, 6760 pp.~390--393, (1999).
\href{http://dx.doi.org/10.1038/46503}{http://dx.doi.org/10.1038/46503}

\bibitem{Low2009}
R.~Low,
``Learning and testing algorithms for the Clifford group''
\newblock Phys.~Rev.~A \textbf{80}, 052314 (2009).
\href{http://dx.doi.org/10.1103/PhysRevA.80.052314}{http://dx.doi.org/10.1103/PhysRevA.80.052314}


\bibitem{Childs2005}
A.~M.~Childs,
``Secure assisted quantum computation''
\newblock Quantum Info. Comput.\textbf{5}, pp.~456, (2005).
\href{http://dl.acm.org/citation.cfm?id=2011674}{http://dl.acm.org/citation.cfm?id=2011674}



\bibitem{Beigi:2010}
S.~Beigi and P.W.~Shor,
``C3, semi-clifford and generalized semi-clifford operations''
\newblock Quantum Info. Comput.\textbf{10}, 1 pp.~41--59, (2010).
\href{http://dx.doi.org/10.1137/S0097539797324886}{http://dx.doi.org/10.1137/S0097539797324886}

\bibitem{Zeng2008}
Zeng, Bei and Chen, Xie and Chuang, Isaac,
``Semi-Clifford operations, structure of Ck hierarchy, and gate complexity for fault-tolerant quantum computation'',
\newblock Phys.~Rev.~A \textbf{77}, 042313 (2008).
\href{http://dx.doi.org/10.1103/PhysRevA.77.042313}{http://dx.doi.org/10.1103/PhysRevA.77.042313}

\bibitem{Gross:2008}
D.~Gross and M.~van den Nest,
``The LU-LC conjecture, diagonal local operations and quadratic forms over GF(2)''
\newblock Quantum Info. Comput.\textbf{8}, 1 pp.~263--281, (2008).
\href{http://portal.acm.org/citation.cfm?id=2011766}{http://portal.acm.org/citation.cfm?id=2011766}



\bibitem{Eastin2009}
B.~Eastin and E.~Knill,
``Restrictions on Transversal Encoded Quantum Gate Sets''
\newblock Phys.~Rev.~Lett. \textbf{102}, 110502, (2009).
\href{http://dx.doi.org/10.1103/PhysRevLett.102.110502}{http://dx.doi.org/10.1103/PhysRevLett.102.110502}

\bibitem{Zeng2007}
B.~Zeng, H.~Chung, A.~Cross and I.~Chuang,
``Local unitary versus local Clifford equivalence of stabilizer and graph states'',
\newblock Phys.~Rev.~A \textbf{75}, 032325 (2007).
\href{http://link.aps.org/doi/10.1103/PhysRevA.75.032325}{http://link.aps.org/doi/10.1103/PhysRevA.75.032325}


\bibitem{BravyiKitaev:2005}
S.~Bravyi and A.~Kitaev,
``Universal quantum computation with ideal Clifford gates and noisy ancillas'',
\newblock Phys.~Rev.~A \textbf{71}, 022316 (2005).
\href{http://dx.doi.org/10.1103/PhysRevA.71.022316}{http://dx.doi.org/10.1103/PhysRevA.71.022316}

\bibitem{Knill05}
E.~Knill,
``Quantum Computing with Realistically Noisy Devices''
\newblock Nature \textbf{434}, pp.~39--44, (2005).
\href{http://dx.doi.org/10.1038/nature03350}{http://dx.doi.org/10.1038/nature03350}


 \bibitem{WvDMH:2010}
W.~van~Dam and M.~Howard,
``Noise thresholds for higher-dimensional systems using the discrete Wigner function"
\newblock Phys.~Rev.~A. \textbf{83}, 032310, (2011).
\href{http://dx.doi.org/10.1103/PhysRevA.83.032310}{http://dx.doi.org/10.1103/PhysRevA.83.032310}

\bibitem{Campbell:arxiv12}
E.~T.~Campbell, H.~Anwar and D.~E.~Browne
 ``Magic state distillation in all prime dimensions using quantum Reed-Muller codes'',
arXiv:1205.3104v1, (2012).


\bibitem{Gottesman1999a}
 D.~Gottesman,
``Fault-Tolerant Quantum Computation with Higher-Dimensional Systems'',
in \emph{Quantum Computing and Quantum Communications}, (editor: Colin Williams),
\emph{Lecture Notes in Computer Science,} Volume 1509, (Springer Berlin / Heidelberg, 1999), pp.~302--313.
\href{http://dx.doi.org/10.1016/S0960-0779(98)00218-5}{http://dx.doi.org/10.1016/S0960-0779(98)00218-5}

\bibitem{Appleby:arxiv09}
 D.~M.~Appleby,
 ``Properties of the extended Clifford group with applications to SIC-POVMs and MUBs'',
arXiv:quant-ph/0909.5233, (2009).

 \bibitem{Zhu:2010}
H.~Zhu,
``SIC POVMs and Clifford groups in prime dimensions''
\newblock J.~Phys.~A \textbf{43}, 305305, (2010).
\href{http://dx.doi.org/10.1088/1751-8113/43/30/305305}{http://dx.doi.org/10.1088/1751-8113/43/30/305305}


 \bibitem{Rennes:2004}
J.~M.~Renes, R.~Blume-Kohout, A.~J.~Scott and C.~M.~Caves,
``Symmetric informationally complete quantum measurements''
\newblock J.~Math.~Phys. \textbf{45}, 2171, (2004).
\href{http://link.aip.org/link/doi/10.1063/1.1737053}{http://link.aip.org/link/doi/10.1063/1.1737053}

\bibitem{Quotation}
Recall the aphorism ``All primes are odd except two, which is the oddest prime of all''.


\bibitem{Zauner:1999}
Gerhard Zauner,
``Grundzüge einer nichtkommutativen Designtheorie'' or ``Foundations of a non-commutative Design Theory'', Ph.~D.~ Thesis, University of Vienna (1999). English translation available at
\newblock IJQI \textbf{9}, 1, pp.~445-507. (2011).
\href{http://dx.doi.org/10.1142/S0219749911006776}{http://dx.doi.org/10.1142/S0219749911006776}


 \bibitem{Cormick:2006}
 C.~Cormick,  E.~F.~Galv\~ao, D.~Gottesman, J.~Pablo~Paz, and A.~O.~Pittenger ,
``Classicality in discrete Wigner functions"
\newblock Phys.~Rev.~A. \textbf{73}, 012301, (2006).
\href{http://dx.doi.org/10.1103/PhysRevA.73.012301}{http://dx.doi.org/10.1103/PhysRevA.73.012301}

\bibitem{Gross:2006}
 D.~Gross,
``Hudson's theorem for finite-dimensional quantum systems''
\newblock J.~Math.~Phys. \textbf{47}, number 12, 122107, (2006).
\href{http://link.aip.org/link/doi/10.1063/1.2393152}{http://link.aip.org/link/doi/10.1063/1.2393152}


\bibitem{Veitch:arxiv12}
V.~Veitch, C.~Ferrie, J.~Emerson
 ``Negative Quasi-Probability Representation is a Necessary Resource for Magic State Distillation'',
Arxiv preprint arXiv:1201.1256, (2012).
\href{http://arxiv.org/abs/1201.1256}{http://arxiv.org/abs/1201.1256}


\bibitem{Negativity}
Note that the definition of negativity used here is essentially that of \cite{Cormick:2006}, and this differs substantially from that given in \cite{Gross:2006,Veitch:arxiv12}. 



\bibitem{Buhrmanetal:2006}
 H.~Buhrman, R.~Cleve, M.~Laurent, N.~Linden, A.~Schrijver and F.~Unger,
``New Limits on Fault-Tolerant Quantum Computation'',
 \emph{Annual IEEE Symposium on Foundations of Computer Science,}
 pp.~411--419, (2006).
\href{http://dx.doi.org/10.1109/FOCS.2006.50}{http://dx.doi.org/10.1109/FOCS.2006.50} 

\bibitem{VirmaniHuelgaPlenio:2005}
S.~Virmani, S~F.~Huelga and M~B.~Plenio,
\newblock Classical simulability, entanglement breaking, and quantum
  computation thresholds,
Phys.~Rev.~A \textbf{71}, 042328 (2005).
\href{http://dx.doi.org/10.1103/PhysRevA.71.042328}{http://dx.doi.org/10.1103/PhysRevA.71.042328}



\bibitem{Plenio2010}
M.~B.~Plenio and S.~Virmani,
``Upper bounds on fault tolerance thresholds of noisy Clifford-based quantum computers''
\newblock New Journal of Physics \textbf{12}, number 3, 033012 , (2010).
\href{http://dx.doi.org/10.1088/1367-2630/12/3/033012}{http://dx.doi.org/10.1088/1367-2630/12/3/033012}



\bibitem{ReichardtMagic09}
B.~Reichardt,
``Quantum universality by state distillation''
\newblock Quantum Inf. Comput. \textbf{9},  pp.~1030--1052 (2009).
\href{http://dx.doi.org/10.1007/s11128-005-7654-8}{http://dx.doi.org/10.1007/s11128-005-7654-8}





%
%
%
%
%






 \bibitem{WvDMH:2009}
W.~van~Dam and M.~Howard,
``Tight Noise Thresholds for Quantum Computation with Perfect Stabilizer Operations"
\newblock Phys.~Rev.~Lett. \textbf{103}, 170504, (2009).
\href{http://dx.doi.org/10.1103/PhysRevLett.103.170504}{http://dx.doi.org/10.1103/PhysRevLett.103.170504}



%

%
%










\bibitem{Byrd:2011}
M.~Byrd, C.~Bishop and Y.~Ou,
``General open-system quantum evolution in terms of affine maps of the polarization vector''
\newblock Phys.~Rev.~A. \textbf{83}, 012301, (2011).
\href{http://dx.doi.org/10.1103/PhysRevA.83.012301}{http://dx.doi.org/10.1103/PhysRevA.83.012301}


\bibitem{NRS:2006}
G.~Nebe, E.~M.~Rains, and N.~J.~A.~Sloane,
\newblock Self-Dual Codes and Invariant Theory (Springer: Berlin, 2006).

\bibitem{NRS:2001}
G.~Nebe, E.~M.~Rains and N.~J.~A.~Sloane,
``The Invariants of the Clifford Groups'',
\newblock Designs, Codes and Cryptography \textbf{24}, 1 99--122 (2001).
\href{http://dx.doi.org/10.1023/A:1011233615437}{http://dx.doi.org/10.1023/A:1011233615437}

\bibitem{CampbellBrowne:2010}
E.~T.~Campbell and D.~E.~Browne,
``Bound States for Magic State Distillation in Fault-Tolerant Quantum Computation"
\newblock Phys.~Rev.~Lett. \textbf{104}, 030503, (2010).
\href{http://dx.doi.org/10.1103/PhysRevLett.104.030503}{http://dx.doi.org/10.1103/PhysRevLett.104.030503}

\end{thebibliography}
\end{document}